\documentclass[12pt, preprint]{aastex}

\shorttitle{A He~I Case-B Recombination Code}
\shortauthors{Porter}

\begin{document}
\title{A He~I Case-B Recombination Code}


\author{R. L. Porter}
\affil{Dept. of Physics and Astronomy, University of Kentucky, Lexington, KY, 40506}
\email{rporter@pa.uky.edu}

\begin{abstract}
Recent calculations of collisionless, Case-B, He~I emissivities were performed by Bauman et al. (2005).  The source code used in the calculation has been freely available online since that paper was published.  A number of changes have been made to simplify the use of the code by third parties.  Here I provide details on how to obtain, compile, and execute the program and interpret the results.  
\end{abstract}


\section{Introduction}
This document serves a simple purpose: to inform interested workers about how to obtain and use the ``$J$-resolved'' computer program discussed in Bauman et al. (2005), hereafter referred to as ``the code.''  There are no new scientific results presented here, and the methods and predictions of the code are unchanged.   

\section{Obtaining the Code}
The code can be obtained at http://www.pa.uky.edu/$\sim$rporter/j-resolved/.  There are two versions available, but the difference is only in the packaging.  The file \verb|heliumRecomb.tar.gz| is convenient for use on Unix and Linux systems, while the file \verb|heliumRecomb.zip| is more convenient for Microsoft Windows machines.

\section{Compiling the Code}

While I have tested the code on several systems, I cannot guarantee that it will compile correctly on any given system.  Nonetheless, the code will likely work on any version of Windows, Unix, Linux, or BSD.  I have made no attempt to compile the code in MacOS.  

Before compiling, you may wish to edit the file \verb|Helium.h| so that the variable \linebreak OUTPUT\_DIR on the first line points to a directory other than where the source files are located.  Now you are ready to compile.  The preferred compiler is the open-source GNU compiler, gcc (available at http://gcc.gnu.org/).  Compile with gcc using the following commands:
\begin{verbatim}
gcc -c *.c
gcc -o EXECUTABLE *.o -lm
\end{verbatim}
where EXECUTABLE is whatever filename you wish to give to your executable \linebreak (i.e., \verb|helium.exe|).
 
The code can also be compiled with Microsoft Visual Studio.  First, create a new project from the file/new project option. In ``Project Types'' select ``Visual C++ Projects.''   Under ``templates'' select ``Win32 Project'' and enter a name for the project (the name of the executable will be the project name with ``.exe'' appended).  Click OK.  Next the Win32 Application Wizard opens.  Click ``application settings.''  Select ``console application'' and ``empty project'' then click on ``finish.''  Now add all the source and header files to the project with the ``project/add existing files'' option.  The debug and optimized versions are referred to as ``debug'' and ``release.''  Compile the source with the ``build'' commands in the ``Build'' menu. 

\section{Executing the Code}
If there were no errors in the previous step, you should now have a working executable.  Next, simply type the executable filename at the command prompt in the directory where your exectuable was created.  Note that some systems will not automatically look for the executable in the current directory.  In this case, the command would be ``./helium.exe''.  The code should respond by printing the following output and then exiting:
\begin{verbatim}
USAGE: executable nmax Te topoff ST-mixing JobName
executable - the executable filename
nmax       - the highest principal quantum number to explicitly consider
Te         - the electron temperature in Kelvin
topoff     - switch that enables (1) or disables (0) recombination above nmax
ST-mixing  - switch that enables (1) or disables (0) singlet-triplet mixing
JobName    - a name for this calculation, will also appear in output filenames
\end{verbatim}
Each of the six items above must be specified and they must be specified in that exact order.  The minimum \verb|nmax| is 15.  Running the code with \verb|nmax| $ > 100$ is not recommended because the code is not well tested beyond that limit.  (See Bauman et al. [2005] for a discussion of convergence with respect to \verb|nmax|.)  If singlet-triplet mixing is disabled, the results will be for the pure $LS$-coupling case.  Here is an example:
\begin{verbatim}
./helium.exe 15 10000 1 1 testrun
\end{verbatim}
This example will produce results for \verb|nmax|=15 at a temperature of 10,000~K, with topoff and $ST$-mixing included.  The name ``testrun'' will be assigned to the job.  

\section{Interpreting the Results}

The above example will produce the following output files:
\begin{verbatim}
testrun_15_Output.txt
testrun_15_Results.txt
testrun_15_Results_Ryan.txt
testrun_15_Levels_st.txt
testrun_15_Thrshld_X_Scts.txt
testrun_15_Solution_st.txt
testrun_15_Vector_st.txt
testrun_15_OscStrMethod_st.txt
\end{verbatim}
The first two parts of each output filename are given by the \verb|JobName| and \verb|nmax| parameters.  The last part of the filename (before the ``.txt'') describes the file contents as shown in Table~1.

\begin{deluxetable}{ll}
\tabletypesize{\scriptsize}
\tablecaption{Output filenames and contents.}{}
\tablewidth{0pt}
\tablehead
{
  \colhead{Last Part of Filename}     & 
  \colhead{Contents of File}         
}
\startdata
Output & Basic sanity checks and various data; this file can be quite cryptic. \\
Results & $J$-resolved wavelengths, emissivitites, oscillator strengths,\\
 & \ \ \ transition probabilities, occupation numbers, etc. \\
Results\_Ryan & Same as Results but $nLS$-resolved for comparison with \\
 & \ \ \ the He~I model in Cloudy (see Porter et al. 2005, 2007). \\
Levels\_st & Various information about levels.  The ``st'' in the filename \\
 & \ \ \ will be ``ls'' if \verb|ST-mixing| is disabled. \\
Thrshld\_X\_Scts & Threshold photoionization cross-sections from various sources. \\
Solution\_st & The solved vector of level populations (in atomic units). \\
Vector\_st & The initial vector of recombination rates (in atomic units). \\
OscStrMethod\_st & The sources of oscillator strengths for each transition. \\
\enddata
\end{deluxetable}

\section{Referencing Use of the Code}

The primary developer of the code was R. P. Bauman.  Any publication that uses the results of the code should include a reference to Bauman et al. (2005).  For reproducibility, the input parameters should also be reported.

\section{Acknowledgments}

I thank Wan Yan Wong and Leticia Mart\'in Hern\'andez for informing me of their difficulties using the original version of the code.  I also thank University of Kentucky Computer Science graduate students Aarthi Krishnamurthy and Kausalya Madhusudhanan for helping to remedy the situation.  This work was supported by NSF through AST-0607028 and by NASA through NNG05GG04G.


\begin{thebibliography}{}
\bibitem[Bauman et al.(2005)]{bauman05}      Bauman, R. P., Porter, R. L., Ferland, G. J., \& MacAdam, K. B. 2005, ApJ, 628, 541
\bibitem[Porter et al.(2005)]{PBFM}          Porter, R. L., Bauman, R. P., Ferland, G. J., \& MacAdam, K. B. 2005, ApJ, 622L, 73
\bibitem[Porter et al.(2007)]{porter07}      Porter, R. L., Ferland, G. J., \& MacAdam, K. B. 2007, ApJ, 657, 327
\end{thebibliography}
\end{document}